\documentclass[conference,a4paper]{IEEEtran}
\IEEEoverridecommandlockouts
\usepackage{cite}
\usepackage{amsmath,amssymb,amsfonts}
\usepackage{algorithmic}
\usepackage{graphicx}
\usepackage{textcomp}
\usepackage{xcolor}
\usepackage{caption}
\usepackage{mathtools}
\DeclarePairedDelimiter{\ceil}{\lceil}{\rceil}
\usepackage{subcaption}
\def\BibTeX{{\rm B\kern-.05em{\sc i\kern-.025em b}\kern-.08em
    T\kern-.1667em\lower.7ex\hbox{E}\kern-.125emX}}

\IEEEoverridecommandlockouts\IEEEpubid{\makebox[\columnwidth]{ 979-8-3503-1090-0/23/\$31.00~\copyright~2023 IEEE \hfill} \hspace{\columnsep}\makebox[\columnwidth]{ }}

\begin{document}

\title{
Probabilistic On-Demand Charging Scheduling for ISAC-Assisted WRSNs with Multiple Mobile Charging Vehicles}

\author{
	\IEEEauthorblockN{\normalsize Muhammad~Umar~Farooq~Qaisar$^{1}$\IEEEauthorrefmark{1},~\IEEEmembership{Member,~IEEE,}
		Weijie~Yuan$^{1}$\IEEEauthorrefmark{1},~\IEEEmembership{Member,~IEEE,}
		Paolo~Bellavista$^{2}$, \\~\IEEEmembership{Senior Member,~IEEE,} 
		Guangjie~Han$^{3}$,~\IEEEmembership{Fellow,~IEEE,} 
		Rabiu~Sale~Zakariyya$^{1}$,~\IEEEmembership{Member,~IEEE,} and \\
		Adeel~Ahmed$^{4}$,~\IEEEmembership{Student Member,~IEEE,}}
	\IEEEauthorblockA{$^{1}$\,\normalsize Department of Electronic and Electrical Engineering, Southern University of Science and Technology, Shenzhen, China\\
	}
	\IEEEauthorblockA{$^{2}$\,\normalsize Department of Computer Science and Engineering, University of Bologna, Bologna, Italy\\
		}
        \IEEEauthorblockA{$^{3}$\,\normalsize Department of Internet of Things Engineering, Hohai University, Changzhou, China\\
		}
        \IEEEauthorblockA{$^{4}$\,\normalsize School of Computer Science and Technology, University of Science and Technology of China, Hefei, China\\
		}
		\IEEEauthorblockA{\normalsize \IEEEauthorrefmark{1}Corresponding authors: Weijie Yuan and Muhammad Umar Farooq Qaisar, Email:\{yuanwj, muhammad\}@sustech.edu.cn}}

\maketitle

\begin{abstract}
The internet of things (IoT) based wireless sensor networks (WSNs) face an energy shortage challenge that could be overcome by the novel wireless power transfer (WPT) technology. The combination of WSNs and WPT is known as wireless rechargeable sensor networks (WRSNs), with the charging efficiency and charging scheduling being the primary concerns. Therefore, this paper proposes a probabilistic on-demand charging scheduling for integrated sensing and communication (ISAC)-assisted WRSNs with multiple mobile charging vehicles (MCVs) that addresses three parts. First, it considers the four attributes with their probability distributions to balance the charging load on each MCV. The distributions are residual energy of charging node, distance from MCV to charging node, degree of charging node, and charging node betweenness centrality. Second, it considers the efficient charging factor strategy to partially charge network nodes. Finally, it employs the ISAC concept to efficiently utilize the wireless resources to reduce the traveling cost of each MCV and to avoid the charging conflicts between them. The simulation results show that the proposed protocol outperforms cutting-edge protocols in terms of energy usage efficiency, charging delay, survival rate, and travel distance. 
\end{abstract}

\begin{IEEEkeywords}
Wireless rechargeable sensor networks, on-demand, partial charging, ISAC, mobile charging vehicles.
\end{IEEEkeywords}

\section{Introduction}
The internet of things (IoT) based wireless sensor networks (WSNs) have seen significant growth in a broad range of applications over the last decades, including industrial automation, military applications, smart cities, environmental monitoring, and healthcare. A WSN is composed of numerous battery-operated sensor nodes that continuously monitor the state of the environment and send the sensed data to a sink through single or multi-hop communication. A low battery level in sensor nodes can reduce the lifetime of WSNs \cite{1, R5}. Due to the small batteries that power the sensor nodes, energy consumption becomes a significant challenge for WSN applications. In recent years, numerous research studies have been conducted to extend network lifetime. These studies are classified into two categories: energy replenishment \cite{36} and energy conservation \cite{5}.

\textit{Wireless Power Transfer} (WPT) technology has made it possible to effectively replenish the energy in wireless rechargeable sensor networks (WRSNs). A WRSN generally consists of three components: a sink that also serves as a depot for mobile charging vehicles (MCVs), single or multiple MCVs, and sensor nodes with rechargeable batteries that allow MCVs to recharge them by transmitting wireless signals \cite{R7}. The way the sensor nodes are recharged allows the WRSNs to operate continuously in an efficient manner \cite{8}. 


The MCVs generally use one of two fundamental charging scheduling strategies: on-demand charging \cite{9} and periodic charging \cite{12}. In periodic charging, an MCV travels the network according to a predetermined and known schedule. Unfortunately, due to the dynamic energy depletion rate of the nodes, this charging schedule is not ideal. On the contrary, on-demand charging appears to be more realistic because it makes real-time decisions based on the energy requirements of the sensor nodes. Therefore, it can handle situations where the node\textquotesingle s energy depletion rate is highly unpredictable. Moreover, charging strategies generally follow either full charging or partial charging models. In a full-charging model, the sensor nodes receive a full recharge to their battery capacity, resulting in significant charging delays. In contrast, the partial charging model allows for more sensor nodes to be recharged. However, in a charging process, the traveling time and conflicting between multiple MCVs must also be addressed properly.


To address the aforesaid issue, we intend to employ the \textit{Integrated Sensing and Communication} (ISAC) technique \cite{14}, which combines sensing and communication functionality to efficiently use wireless resources, realize wide area environment sensing, and even to pursue mutual benefits. Therefore, ISAC can improve charging efficiency and reduce travel time of MCVs by leveraging the benefits of wireless signal \cite{16}. A charging node can be in a priority queue of multiple MCVs but with different priorities. Thus, whenever an MCV enters the sensing range of a prioritized charging node, the node will send the ISAC signal to the MCV, receive the echo from the MCV via wireless transmission, analyze it, and then communicate with the sink node to update the priority queues of other MCVs. This mechanism significantly reduces travel time and avoids conflicts when multiple MCVs attempt to overcharge a node in a single request.

The aforementioned studies have provided convincing motivations to address the issues of deploying multiple MCVs and developing an effective on-demand charging strategy for sensor nodes, as well as integrating the ISAC concept with WRSNs to optimize network stability. Therefore, this paper proposes a probabilistic on-demand charging scheduling for ISAC-assisted WRSNs with multiple mobile charging vehicles (\textit{Poised}). The contribution of our paper lies in several unique aspects. Firstly, it offers a balanced charging load strategy on each MCV priority queue by utilizing four attributes with their probability distributions, namely the residual energy of the charging node, the distance from the MCV to the charging node, the degree of a charging node, and the charging node betweenness centrality. These attributes help achieve charging efficiency and network lifetime. Secondly, it employs a charging factor strategy for each MCV queue, which partially charges all nodes while further improving charging efficiency and coverage. Lastly, the integration of the ISAC concept with WRSNs efficiently utilizes wireless resources to reduce travel time and avoid conflicts between multiple MCVs overcharging the same node. The primary goal of our paper is to provide a well-balanced charging strategy for multiple MCVs with an efficient on-demand charging strategy to optimize charging efficiency. Additionally, we integrate sensing and communication techniques to reduce MCV travel time within the network, thereby improving network stability. The unique aspects and contributions of our approach make our proposal an innovative and effective solution to the identified issues.


\section{Related Work}
This section provides a brief overview of several studies on WRSNs energy replenishment that are relevant to our work.

The authors of \cite{18} developed a charging method that clusters the energy requirements of nodes in order to equally distribute the charging load across the MCVs. It reduces the charging delay to some extent by increasing the number of recharged nodes. In \cite{19} and \cite{20}, the authors introduced charging scheduling approaches for the problem of minimizing the longest delay. To prevent a sensor node from being charged by two or more MCVs at once, they sought to identify a closed charging tour for each MCV. However, the charging load across the MCVs were unbalanced due to their methodology. 
The authors of \cite{9} introduced a distributed mobile charging protocol to schedule multiple MCVs in dense WRSNs. They focused on an on-demand partial charging strategy and applied a game theory technique to address the multiple charging problem, which resulted in repetitive games played by the MCVs. Their approach reduced charging delay to some extent by improving charging coverage. The authors of \cite{27} addressed the problem of multiple MCVs coordination, which consists of scheduling multiple chargers and optimizing travel time with the goal of reducing the overall energy usage of MCVs by modifying their mobility speed and charging time. The authors of \cite{29} introduced an uneven cluster-based mobile charging approach that allocates nodes into clusters and then focuses on the charging schedule for each MCV while taking remaining energy and sensor node distance into account. Despite this, their approach has a lower charging efficiency. 
The authors of \cite{8} introduced a charging scheduling approach based on fuzzy logic with multiple MCVs. Their approach divides the network to evenly distribute the charging load of sensor nodes on each MCV, and it also determines a dynamic charging threshold for each node depending on their respective energy usage rates. Furthermore, for each MCV, the next node to be charged is determined by combining fuzzy logic and multi-metric inputs. However, due to an inefficient multi-metric strategy, their approach falls short in the efficient selection of the next node to be charged. The authors of \cite{31} introduced an approach for reducing charging delays by charging sensor nodes with multiple MCVs. Unlike prior studies, the travel trajectories of MCVs were predetermined, and their speed was varied.

The aforementioned approaches demonstrated the feasibility and acceptability of charging scheduling strategies in WRSNs. However, these approaches lacked a combined focus on an effective multi-metric strategy for balancing the charging load on the MCVs and an efficient charging factor strategy for each MCV to partially charge the network nodes. Additionally, they did not effectively address the issue of reducing travel time for multiple MCVs and their charging conflicts in the network. Therefore, this paper attempts to address the aforesaid issues by employing an effective multi-metric and efficient charging factor strategy. It also applies ISAC in WRSNs to reduce travel time and avoid charging conflicts through sensing and communication tasks between charging nodes and MCVs.


\section{The Proposed Protocol} \label{TPP}
This paper proposes probabilistic on-demand charging for ISAC-assisted WRSNs with multiple mobile charging vehicles (\textit{Poised}), and it goes into detail about effective strategies for balancing the charging load on each MCV. It also partially charges all of the nodes in their queues in an efficient manner to improve charging coverage and efficiency. Furthermore, it integrates the ISAC concept by utilizing wireless resources to reduce the traveling time of each MCV in the network.

\subsection{Network Model and Initialization} \label{NMA}
A standard wireless rechargeable sensor network is assumed to consist of a set of randomly deployed sensor nodes, $\mathcal{N} = \{{Sn_{0}, Sn_{1}, Sn_{2},..., Sn_{r}}\}$ and a set of mobile charging vehicles (MCVs), $\mathcal{V} = \{{Cv_{1}, Cv_{2}, Cv_{3},..., Cv_{m}}\}$, with $m = |\mathcal{V}|$ in a two-dimensional region. When the energy level of the sensor nodes ($Sn_{i}$) falls below a certain level called threshold ($\mathcal{E}_{th}$), charging requests are sent to the sink node ($Sn_{b}$). The sink is positioned in the center of the region and also serves as a depot for the MCVs. In this work, the initial position of the MCVs is adapted from \cite{9}. The Eq. \eqref{MCV1} is employed in particular to determine the coordinates $(x_{i}, y_{i})$ of MCVs initial position in $k$ different regions.

\begin{equation} \label{MCV1}  \small 
	(x_{i},y_{i}) = \left(\frac{\mathcal{C}_{c}}{2}\cos{\left(\frac{\pi}{m}(2j-1)\right)}, \frac{\mathcal{C}_{c}}{2}\sin{\left(\frac{\pi}{m}(2j-1)\right)}\right)
\end{equation}

Here, $\mathcal{C}_{c}$ is the radius of circumscribed circle of a two-dimensional region. The MCV recharges one sensor node ($Sn_{i}$) at a time. It should be noted that an MCV has a substantially higher energy capacity than a sensor node ($Sn_{i}$). Table \ref{Notations} provides a summary of the notations used in this work.

\begin{table}[h]
	\centering
	\caption{Notations}
	\label{Notations}
	\begin{tabular}
		{cp{7cm}}
		\hline
		\textbf{Notation} & \textbf{Definition} \\
		\hline
		$\mathcal{N}$  & $\mathcal{N} = \{{Sn_{0}, Sn_{1}, Sn_{2},..., Sn_{r}}\}; Sn_{i} \in \mathcal{N}$ is a sensor node$; r$ is the size of $\mathcal{N}$.\\
		$\mathcal{N}_{i}$ & The neighboring nodes of $Sn_{i}$; $r_{i}$ is the size of $\mathcal{N}_{i}$. \\
		$\mathcal{V}$  & $\mathcal{V} = \{{Cv_{1}, Cv_{2}, Cv_{3}, ..., Cv_{m}}\}; Cv_{j} \in \mathcal{V}$ is a mobile charging vehicle$; m$ is the size of $\mathcal{V}$.\\
  	    $Sn_{s}$ & The source node.   \\
		$Sn_{b}$ & The sink.   \\
		$\mathcal{R}_{s}$ & The sensing range of node $Sn_{i}$. \\
		$\mathcal{R}_{c}$ & The communication range of node $Sn_{i}$. \\
            $\mathcal{E}_{th}$ & The residual energy threshold of node $Sn_{i}$. \\
            $\mathcal{CE}_{th}$ & The minimum working energy threshold of MCV $Cv_{j}$. \\
		$\Phi_{i}^\downarrow$ & The residual energy of node below threshold $\mathcal{E}_{th}$. \\
		$\zeta_{i}$ & The distance from MCV $Cv_{j}$ to node $Sn_{i}$. \\
		$\eta_{i}$ & The degree of node $Sn_{i}$. \\
            $\ss_{i}$ & The node $Sn_{i}$ betweenness centrality. \\
            $\mathcal{C}_{c}$ & The radius of circumscribed circle. \\ 
		$e$ & The \textit{Euler\textquotesingle s Constant}. Its approximate value is 2.71828. \\
		$\pi$ & The default value of Pi is approximately equal to 3.14159. \\
		\hline
	\end{tabular}
\end{table}

\subsection{Charging Queue Metric} \label{CQM}
This section describes the balanced charging load strategy for each MCV queue, 
which considers the four important attributes and their respective probability distributions to prioritize the charging queue metric. These attributes are the residual energy of charging node Eq. \eqref{RD2}, distance from MCV to charging node Eq. \eqref{DMC3}, degree of a charging node Eq. \eqref{DCN3}, and charging node betweenness centrality Eq. \eqref{CNB3}. The charging queue metric will prioritize the charging nodes with highest value. Therefore, each MCV will have a different sequence of prioritized charging nodes due to their different locations in the network as described in Section \ref{NMA}. 

\subsubsection{Residual Energy of Charging Node} \label{RECN}
The purpose of this distribution is to prioritize the sensor nodes which have the lower residual energy among all the charging sensor nodes. Here, we express the residual energy of a charging sensor node $(i.e., \forall\, Sn_{i} \in \mathcal{N})$ as a vector, $\Phi_{i}^\downarrow= (\Phi_{1}^\downarrow,\Phi_{2}^\downarrow,\Phi_{3}^\downarrow,...,\Phi_{r}^\downarrow)$, where $\Phi_{i}^\downarrow$ is the residual energy of a charging sensor node below the residual energy threshold $\mathcal{E}_{th}$. The residual energy vector is normalized into $\bar{\Phi}_{i}^\downarrow= (\bar{\Phi}_{1}^\downarrow,\bar{\Phi}_{2}^\downarrow,\bar{\Phi}_{3}^\downarrow,...,\bar{\Phi}_{r}^\downarrow)$ between $[0-1]$ using Eq. \eqref{RD1}. Based on it by curve fitting the normalized vector, we obtain the probability distribution function using Eq. \eqref{RD2}, $\Tilde{\Phi}_{i}= (\Tilde{\Phi}_{1}^\downarrow,\Tilde{\Phi}_{2}^\downarrow,\Tilde{\Phi}_{3}^\downarrow,...,\Tilde{\Phi}_{r}^\downarrow)$. 
Furthermore, we included a weighted factor $\lambda_{\Phi}$ to increase the impact of the distribution, where $\lambda_{\Phi}\ge2$ is set by default. Nodes with lower residual energy will have higher charging priority. 

\begin{equation} \label{RD1}
\bar{\Phi}_{i}^\downarrow = \frac{\Phi_{i}^\downarrow}{\mathcal{E}_{th}} \quad\forall \, Sn_{i} \in \mathcal{N}
\end{equation}

\begin{equation} \label{RD2}
\begin{array}{c@{\,}c}
\tilde{\Phi}_{i}^\downarrow=\begin{cases}
\alpha + \beta * \mathrm{e}^{\left(\gamma * \left(\bar{\Phi}_{i}^\downarrow \right)^{\lambda_{\Phi}}\right)} \\
\alpha=4.1997;\quad \beta=-3.16759; \\ \gamma =0.27897; \quad  \lambda_{\Phi}=2; \\

\end{cases} \,\forall \, Sn_{i} \in \mathcal{N}
\end{array}
\end{equation}

\subsubsection{Distance from MCV to Charging Node} \label{DMCN}
The purpose of this distribution is to prioritize the sensor nodes which have the shortest distance among all the charging sensor nodes to MCV. Here, we use the \textit{Euclidean Distance} to express the distance from each MCV $(i.e., \forall\, Cv_{j} \in \mathcal{V})$ to charging sensor node $(i.e., \forall\, Sn_{i} \in \mathcal{N})$ as a vector, $\zeta_{v_{j},i}= (\zeta_{v_{j},1},\zeta_{v_{j},2},\zeta_{v_{j},3},...,\zeta_{v_{j},r})$, which is obtained in Eq. \eqref{DMC1}. The distance vector is normalized into $\Bar{\zeta}_{v_{j},i}= (\Bar{\zeta}_{v_{j},1},\Bar{\zeta}_{v_{j},2},\Bar{\zeta}_{v_{j},3},...,\Bar{\zeta}_{v_{j},r})$ between $[0-1]$ using Eq. \eqref{DMC2}. The probability distribution function is used to curve fit the normalized vector, as stated in Eq. \eqref{DMC3}, $\Tilde{\zeta}_{v_{j},i}= (\Tilde{\zeta}_{v_{j},1},\Tilde{\zeta}_{v_{j},2},\Tilde{\zeta}_{v_{j},3},...,\Tilde{\zeta}_{v_{j},r})$.
Furthermore, we included a weighted factor $\lambda_{\zeta}$ to increase the impact of the distribution, where $\lambda_{\zeta}\ge1$ is set by default. The shorter the distance to MCV will have higher priority of a charging sensor node.


\begin{equation} \label{DMC1}
	\zeta_{v_{j},i} = \sqrt{(x_{j}-x_{i})^2+(y_{j}-y_{i})^2}
\end{equation}

\begin{equation} \label{DMC2}
\bar{\zeta}_{v_{j},i} = \frac{\zeta_{v_{j},i}}{\zeta_{v_{j},i} + \mathcal{R}_{c}} \quad\forall \, Sn_{i} \in \mathcal{N}
\end{equation}

\begin{equation} \label{DMC3}
\begin{array}{c@{\,}c}
\tilde{\zeta}_{v_{j},i}=\begin{cases}
\alpha + \beta * \mathrm{e}^{\left(\frac{\left(\left(\bar{\zeta}_{v_{j},i}\right)^{\lambda_{\zeta}}-\gamma\right)}{\mu}\right)} \\
\alpha=1.83283;\quad \beta=-0.69354; \\ \gamma=-0.24143; \quad \mu =1.28078; \\  \lambda_{\zeta}=1; \\

\end{cases} \,\forall \, Sn_{i} \in \mathcal{N}
\end{array}
\end{equation}

\subsubsection{Degree of Charging Node} \label{DCN}
The purpose of this distribution is to prioritize the sensor nodes which have maximum number of neighbors among all the charging sensor nodes. Here, we express the degree of charging sensor node $(i.e., \forall\, Sn_{i} \in \mathcal{N})$ as a vector, $\eta_{i}= (\eta_{1},\eta_{2},\eta_{3},...,\eta_{r})$, which is obtained in Eq. \eqref{DCN1}. The degree vector is normalized into $\Bar{\eta}_{i}= (\Bar{\eta}_{1},\Bar{\eta}_{2},\Bar{\eta}_{3},...,\Bar{\eta}_{r})$ between $[0-1]$ using Eq. \eqref{DCN2}. 
The normalized vector is curve fitted with the probability distribution function, as indicated in Eq. \eqref{DCN3}, $\Tilde{\eta}_{i}= (\Tilde{\eta}_{1},\Tilde{\eta}_{2},\Tilde{\eta}_{3},...,\Tilde{\eta}_{r})$.
Furthermore, we included a weighted factor $\lambda_{\eta}$ to increase the impact of the distribution, where $\lambda_{\eta}\ge1$ is set by default. The charging sensor node with the maximum number of neighboring nodes will have a higher priority.

\begin{equation} \label{DCN1}
\eta_{i} = \sum_{v=1}^{r_{i}}{Sn_{i,v}} \\
\end{equation}

\begin{equation} \label{DCN2}
\bar{\eta}_{i} = \frac{\eta_{i}}{\max\{\eta_{i}\vert Sn_{i}\in\mathcal{N}\}} \quad\forall \, Sn_{i} \in \mathcal{N}
\end{equation}

\begin{equation} \label{DCN3}
\begin{array}{c@{\,}c}
\tilde{\eta}_{i}=\begin{cases}
\alpha + \beta * \left(\frac{\left(\mathrm{e}^{\left(\gamma * \left(\bar{\eta}_{i}\right)^{\lambda_{\eta}}\right)}-\mu\right)}{\gamma}\right) \\
\alpha=0.02098;\quad \beta=1.29332; \\ \gamma=-0.4591; \quad \mu=0.978; \\ \lambda_{\eta}=1;
\end{cases} \,\forall \, Sn_{i} \in \mathbb{N}
\end{array}
\end{equation}

\subsubsection{Charging Node Betweenness Centrality} \label{CNBC}
The purpose of this distribution is to prioritize the sensor nodes, which act as a bridge between the source node and sink node most of the time. Here, we express the betweenness centrality of a charging sensor node $(i.e., \forall\, Sn_{i} \in \mathcal{N})$ as a vector, $\ss_{i}= (\ss_{1},\ss_{2},\ss_{3},...,\ss_{r})$, which is obtained in Eq. \eqref{CNB1}. The betweenness centrality vector is normalized into $\Bar{\ss}_{i}= (\Bar{\ss}_{1},\Bar{\ss}_{2},\Bar{\ss}_{3},...,\Bar{\ss}_{r})$ between $[0-1]$ using Eq. \eqref{CNB2}. 
The probability distribution function, as described in Eq. \eqref{CNB3}, $\Tilde{\ss}_{i}= (\Tilde{\ss}_{1},\Tilde{\ss}_{2},\Tilde{\ss}_{3},...,\Tilde{\ss}_{r})$, is used to curve fit the normalized vector.
Furthermore, we included a weighted factor $\lambda_{\ss}$ to increase the impact of the distribution, where $\lambda_{\ss}\ge1$ is set by default. The maximum time a node acts as a betweenness will have a higher priority of a charging sensor node.

\begin{equation} \label{CNB1}
\ss_{i} = \sum_{Sn_{s}\neq Sn_{i}\neq Sn_{b}} \frac{\sigma_{Sn_{s}Sn_{b}}(Sn_{i})}{\sigma_{Sn_{s}Sn_{b}}}\\
\end{equation}

\begin{equation} \label{CNB2}
\Bar{\ss}_{i} = \frac{\ss_{i} - \min(\ss_{i})}{\max(\ss_{i}) - \min(\ss_{i})}
\end{equation}

\begin{equation} \label{CNB3}
\begin{array}{c@{\,}c}
\tilde{\ss}_{i}=\begin{cases}
\alpha + \mathrm{e}^{\left(-\mathrm{e}^{\left(-\beta * \left(\left(\Bar{\ss}_{i}\right)^{\lambda_{\ss}}-\gamma\right)\right)}\right)} \\
\alpha=1.343494;\quad \beta=2.88956; \\ \gamma=0.57881; \quad \lambda_{\ss}=1;
\end{cases} \,\forall \, Sn_{i} \in \mathcal{N}
\end{array}
\end{equation}

The purpose of charging queue metric is to use the above four distributions to efficiently prioritize the charging queue of each MCV. 
As discussed in Section \ref{NMA}, the initial position of each MCV is divided into $k$ regions using Eq. \eqref{MCV1}. Therefore, charging sensor nodes will be prioritized differently for each MCV queue due to the distance distribution Eq. \eqref{DMC3} from each MCV to the sensor node. In this work, the sink node is in charge of prioritizing each MCV queue based on the charging queue metric determined in Eq. \eqref{CMQ1}. The charging queue metric obtained the highest priority value based on the average term of four probability distributions.

\begin{equation} \label{CMQ1} 
\tilde{\mathcal{C}}_{qm} = (\tilde{\Phi}_{i}^\downarrow+\tilde{\zeta}_{v_{j},i}+\tilde{\eta}_{i}+\tilde{\ss}_{i})/4 \quad\quad \forall \, Sn_{i} \in \mathcal{N}
\end{equation}

\subsection{Charging Factor Strategy} \label{CFS}
This section focuses on the efficient probabilistic partial charging model by developing a charging factor strategy for all charging nodes prioritized in each MCV queue. For instance, we express a queue of an MCV with charging sensor nodes as $\mathcal{Q}_{Cv_{j}}= \{Sn_{1}, Sn_{2}, Sn_{3},..., Sn_{r}\}$. Here, we consider the residual energy of a charging node distribution Eq. \eqref{RD2} in the queue, where the higher probability of a node is considered to be more critical than the rest of the nodes. 

Therefore, to efficiently evaluate the charging factor strategy and optimize the partial charging in the queue, we first determine the minimum criticality $\mathcal{C}_{min}$ and the maximum criticality $\mathcal{C}_{max}$ of the charging sensor nodes using residual energy priority values, and then obtain the probabilistic weighted factor $\mathcal{P}_{wf}$ in Eq. \eqref{CFS1}.

\begin{equation} \label{CFS1}
\mathcal{P}_{wf} = \frac{\mathcal{C}_{max}(\tilde{\Phi}_{i}^\downarrow) - \mathcal{C}_{min}(\tilde{\Phi}_{i}^\downarrow)}{\mathcal{C}_{max}(\tilde{\Phi}_{i}^\downarrow)}\\
\end{equation}

The charging factor strategy based on Eq. \eqref{CFS1} is obtained in Eq. \eqref{CFS2} to determine the charging factor for each sensor node in the charging queue $\mathcal{Q}_{Cv_{j}}$. 

\begin{equation} \label{CFS2}
\begin{array}{c@{\,}c}
\mathcal{C}_{fs} = \ceil*{\left(\sqrt{(\tilde{\Phi}_{i}^\downarrow)^2 * \mathcal{P}_{wf}} - \partial_{\tilde{\Phi}_{i}^\downarrow} \right) * 100}
\end{array}
\end{equation}

Here, $\partial_{\tilde{\Phi}_{i}^\downarrow}$ is the charging control factor, which is set to 10\% of the residual energy priority Eq. \eqref{RD2} of each queue node. The reason is that the distance between the MCV and each node, as well as the charging time required for each node, must keep in mind to give the MCV a fair chance to successfully partially charge each node in the queue. This helps to improve charging efficiency, coverage, and the network lifetime by preventing nodes from dying during MCV service time. The probabilistic weighted factor, however, is equal to residual energy priority when an MCV only has one sensor node in its request queue.

\subsection{ISAC-based MCV detection}
In this section, we employ the integrated sensing and communication (ISAC) approach, in which the prioritized charging sensor node sends the ISAC signal to the first arriving MCV in its sensing range based on its sensing capability and received echo to analyze which is affected by noise and interference and obtain the distance, and then directly communicate with the sink to prevent other MCVs from visiting the same nodes to overcharge. 
Due to this, it reduces the travel time and the charging delays. The sink then updates the charging queue priority of the other MCVs. The received echo signal $x(t)$ affected by noise and interference is given in Eq. \eqref{ISAC1}.

\begin{equation} \label{ISAC1} 
 x(t) = s(t-\tau) + n(t)
\end{equation} 
where $s(t-\tau)$ is the combination of the original transmitted signal, delayed by the travel time $\tau$ of the echo. $n(t)$ is the additive noise.

Our goal is to detect the presence of the signal $s(t)$ in the noisy received signal $x(t)$. Hence, the sensor node will apply matched filtering to the received signal. The matched filter $h(t)$ is designed to maximize the correlation between the received signal $x(t)$ and the signal $s(t)$ transmitted by the MCV. The output of the matched filter $y(t)$ is given by Eq. \eqref{ISAC2}.

\begin{equation} \label{ISAC2} 
 y(t) = \int x(\tau)h(t-\tau)d\tau 
\end{equation} 
where $h(t)$ is the impulse response of the matched filter. 

To obtain the time delay $\tau$ that maximizes the cross-correlation function Eq. \eqref{ISAC2}, we first rewrite the cross-correlation function as a function of the time delay $\tau$ in Eq. \eqref{ISAC3} and then take the derivative of the cross-correlation function with respect to $\tau$.
\begin{equation} \label{ISAC3} 
y(t) = \int x(\tau)h(t-\tau)d\tau = \int x(t-\tau)h(\tau)d\tau
\end{equation}

According to the concept of waveform matched filter in \cite{37}, the time delay that maximizes the cross-correlation function is given in Eq. \eqref{ISAC4}

\begin{equation} \label{ISAC4}
\tau = \arg\max{\int x(t)s^*(t)d\tau}=\arg\max[x(t)*s^*(t)]
\end{equation}
where $*$ denotes convolution.
In other words, the time delay that maximizes the cross-correlation function is the one that maximizes the convolution of $x(t)$ and $s^*(t)$.

To detect the MCV by the charging sensor node $Sn_{i}$, we estimate the distance between them using a time delay $\tau$ that maximizes the cross-correlation function Eq. \eqref{ISAC4} and speed of light $c = \frac{d}{\tau}$. Hence, the two-way signal distance from charging sensor node to MCV is $2d_{Sn_{i}, Cv_{j}}$. Consequently, it determines the distance in Eq. \eqref{ISAC6}.

\begin{equation} \label{ISAC6}
d_{Sn_{i}, Cv_{j}} = \frac{c * \tau}{2}
\end{equation}

Based on the probabilistic sensing model \cite{36}, if the distance between a sensor node and the MCV $d_{Sn_{i}, Cv_{j}}$, is within the sensing region $\mathcal{R}_{s}$, the sensor node can detect the MCV and communicate with the sink. Subsequently, the sink updates the priority queues of other MCVs by removing that node.

\section{Performance Evaluation and Discussion}
The proposed scheduling protocol, \textit{Poised}, is implemented and evaluated through simulations. Initially, the simulator was developed to assess data routing schemes with both static \cite{1} and mobile \cite{5} sinks using NS3 models. Later on, it was modified to evaluate WRSN techniques \cite{R7}.
The WRSN network is constructed on a square-shaped monitoring area. The sensor nodes are distributed at random, with the sink node in the center of the area. The sink serves as a depot for MCVs to recharge their batteries. The sink is also in charge of scheduling charging requests for each MCV as well as network management. To ensure the accuracy of the results, we took into account the average of $20$ random simulations. Table \ref{SP} contains a summary of the key simulation parameters. 

\begin{table}[h]
	\scriptsize
	\centering
	\caption{Simulation Parameters}
	\label{SP}
	\begin{tabular}{p{4cm}p{3.5cm}}
		\hline
		\textbf{Parameter} & \textbf{Value} \\
		\hline
		Number of nodes  & Varies from $100$ to $500$\\
		Communication range & $50m$ \\
  		Sensing range & $25m$ \\
		Sensor battery capacity & $0.5J$ \\
            Threshold for charging requests & $30\%$ residual energy \\
		MCV battery capacity & $10kJ$ \\
  		The charging rate & $0.05J/s$ \\
		MCV travel speed & $5m/s$ \\
		MCV travel cost & $5J/m$ \\
        MCV mobility model & Random waypoint \\
		\hline
	\end{tabular}
\end{table}

\textit{Poised} is evaluated using a number of parameters that includes: 
i) \textbf{Energy Usage Efficiency} $(\%)$: It is expressed as the ratio of total energy transferred to sensor nodes to total energy transmitted from sink to MCVs. ii) \textbf{Charging Delay} $(s)$: It is expressed as the time it takes the MCVs to complete the energy requirements of the sensor nodes. iii) \textbf{Survival Rate} $(\%)$: It is expressed as the ratio of number of alive sensor nodes to the total number of sensor nodes in the network. iv) \textbf{Travel Distance} $(m)$: It is expressed as the total distance covered by MCV during a single charging tour.

\subsection{Simulation Results}
The proposed protocol \textit{Poised} compares the outcomes with two state-of-the-art protocols: FLCSD \cite{8} and DMCP \cite{9}.

\begin{figure*}[ht]
\centering
\subfloat[Energy Usage Efficiency]{\label{fig: EUENN}\includegraphics[width=.23\linewidth, height=3.4cm]{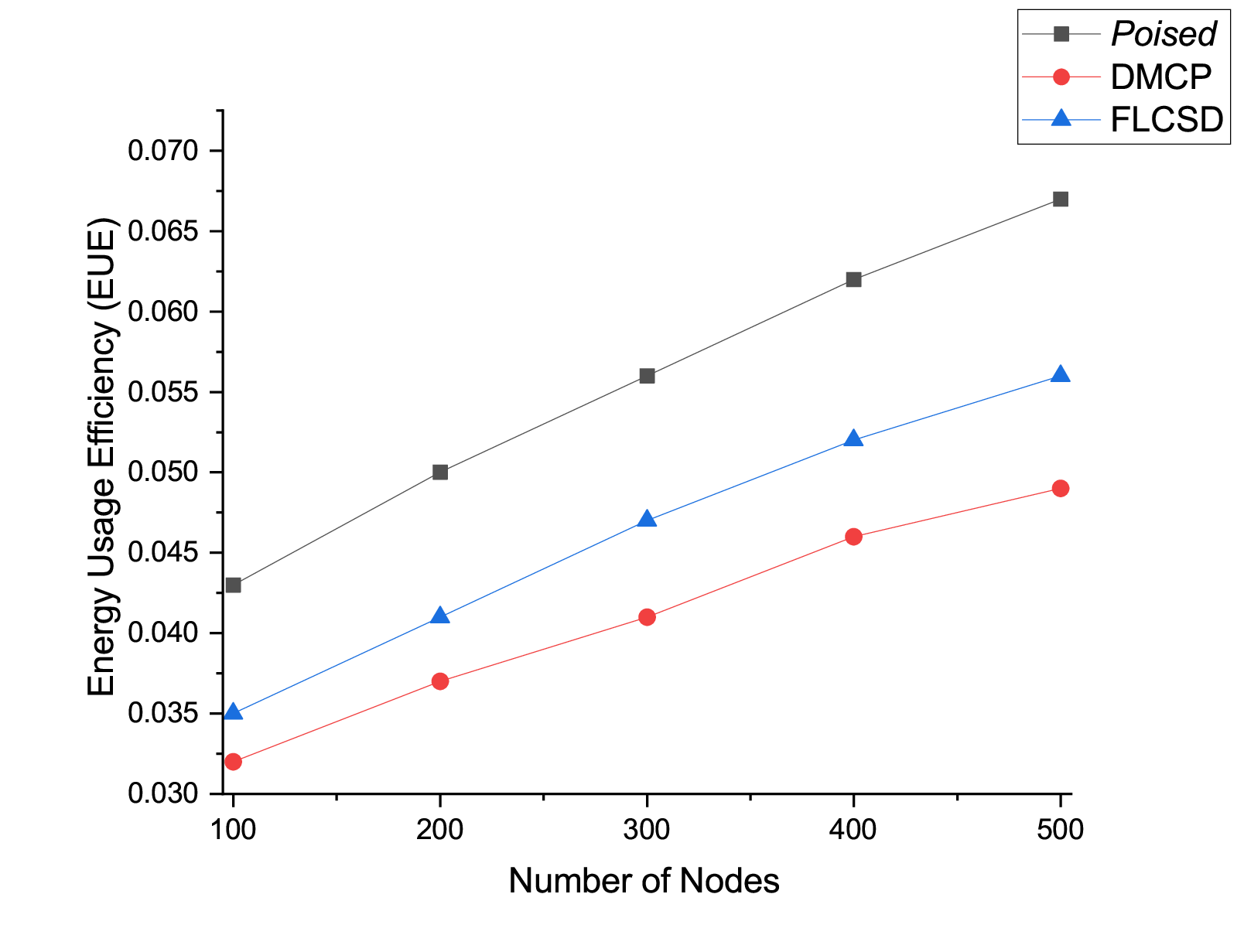}}\quad
\subfloat[Charging Delay]{\label{fig: CDNN}\includegraphics[width=.23\linewidth, height=3.4cm]{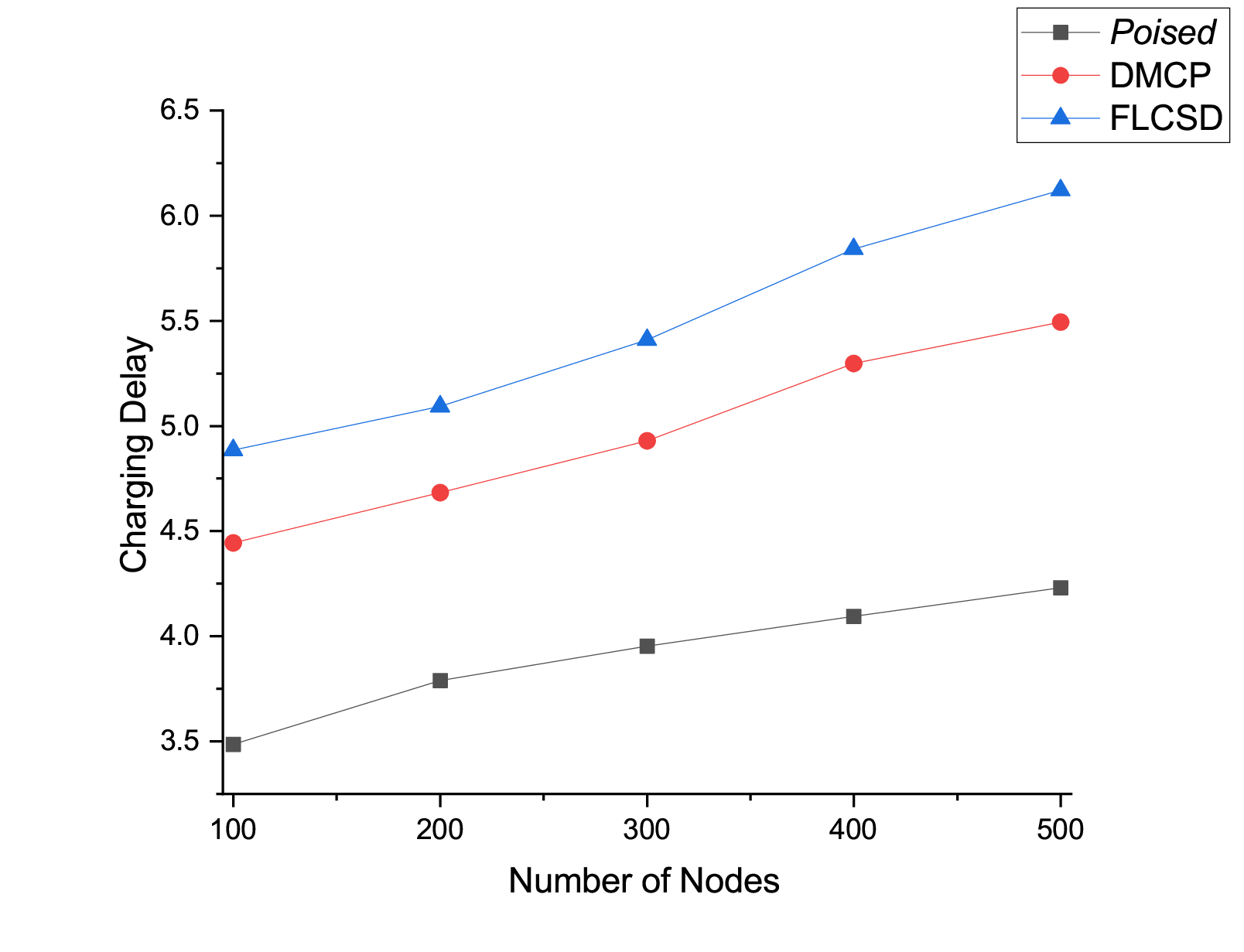}}\quad
\subfloat[Survival Rate]{\label{fig: SRNN}\includegraphics[width=.23\linewidth, height=3.4cm]{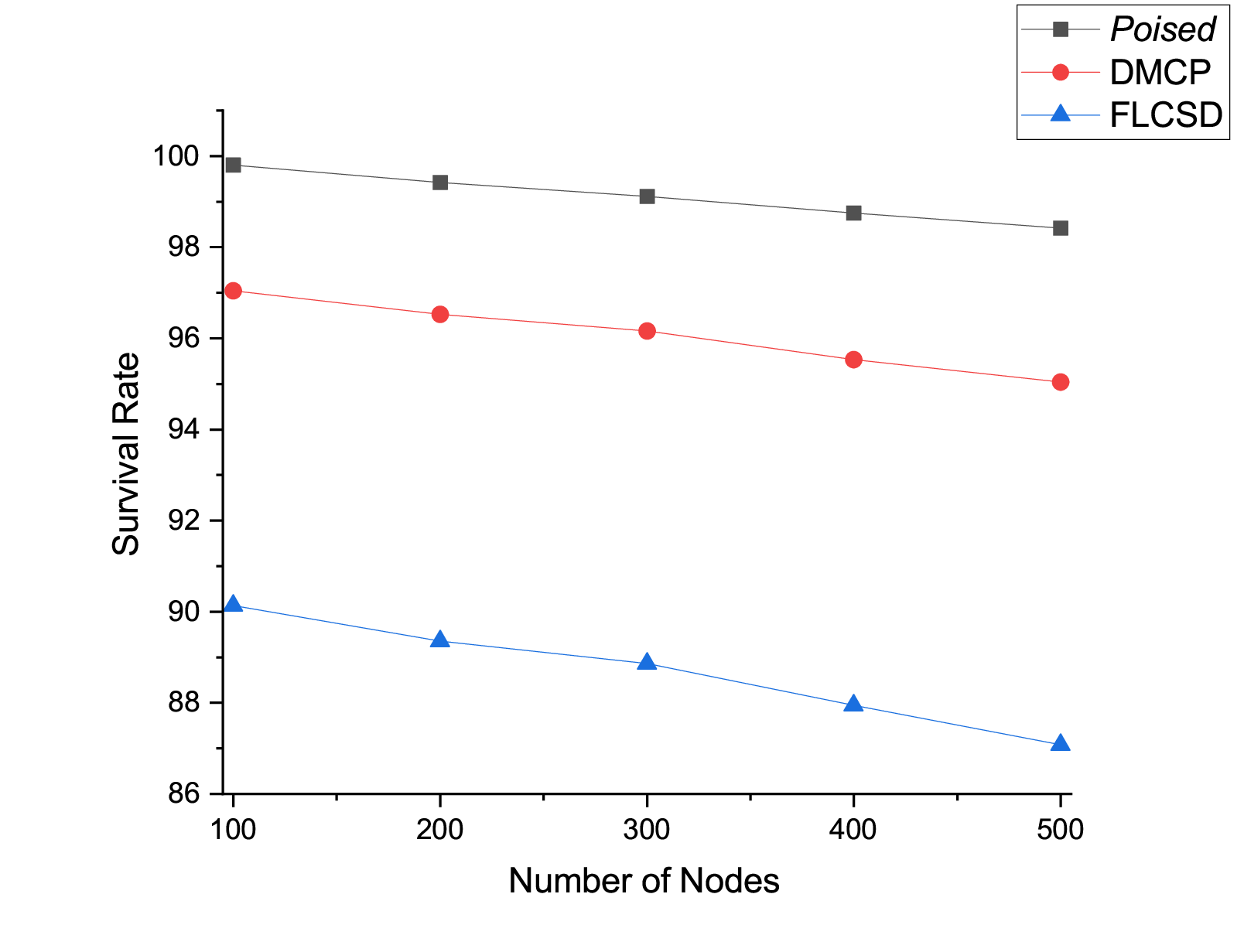}}\quad
\subfloat[Travel Distance]{\label{fig: TDNN}\includegraphics[width=.23\linewidth, height=3.4cm]{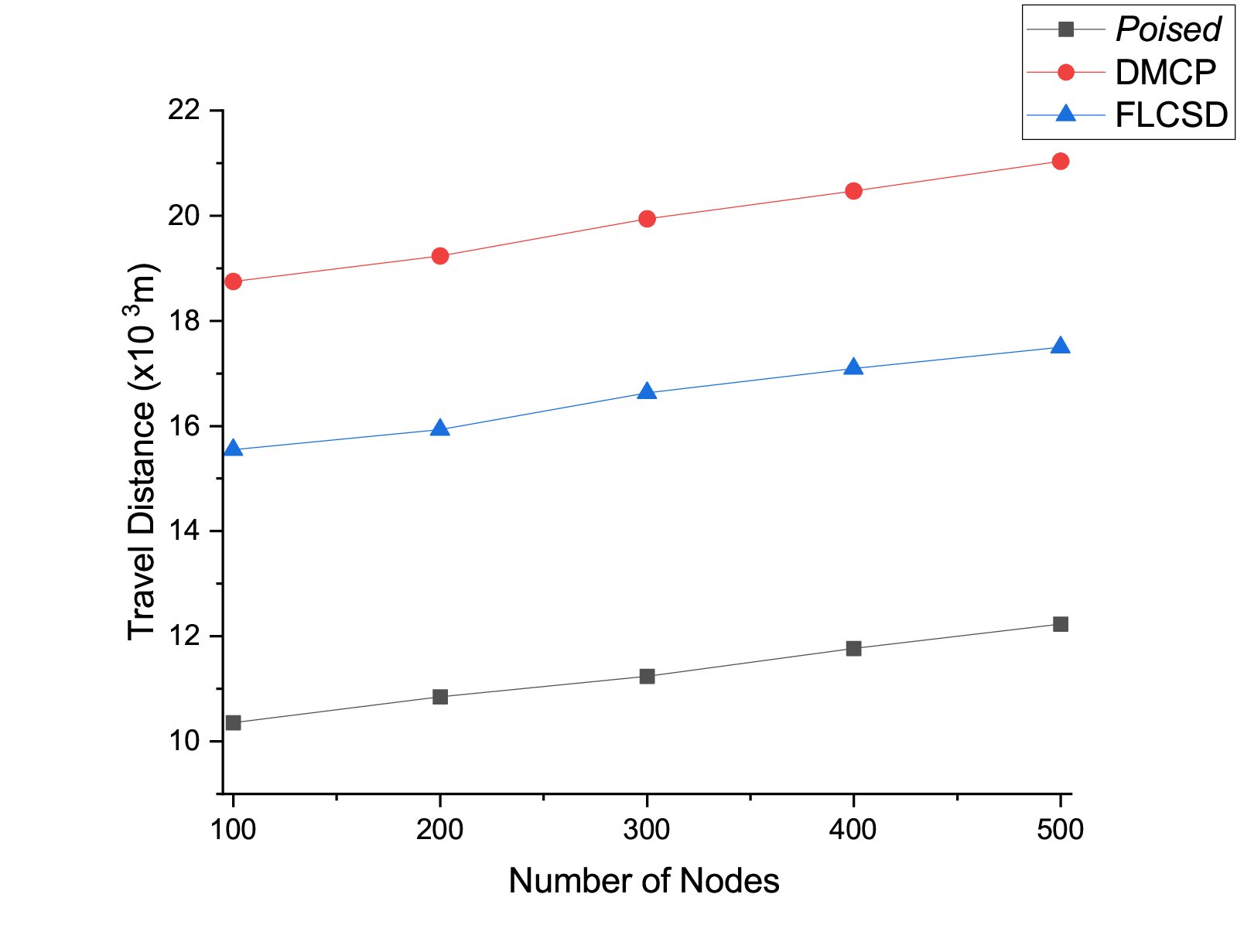}}
\caption{Performance over number of nodes}
\label{fig: PNN}
\end{figure*}

Fig. \ref{fig: EUENN} depicts the results of energy usage efficiency, demonstrating how the result gradually increases with the number of nodes for all protocols. The proposed protocol \textit{Poised} outperforms cutting-edge protocols for the following reasons. First, it employs a well-balanced charging load strategy for each MCV based on the probability distribution of each attribute to fairly prioritize charging nodes in each MCV queue. Second, it provides an effective charging factor strategy in each MCV queue to partially charge all the nodes in the network based on the residual energy of charging node distribution. Finally, it employs the ISAC concept to reduce the travel cost of each MCV in the network. For these reasons, it maximizes the amount of energy transferred to each requested node in the network. FLCSD and DMCP both fell short to consider the efficient charging load for each MCV in the network and to prioritizes the charging queue to partially charge them in order to maximize energy usage efficiency.

Fig. \ref{fig: CDNN} depicts the results of charging delay, demonstrating how the results gradually increases with the number of nodes in the network for all protocols. The proposed protocol \textit{Poised} outperforms cutting-edge protocols because it utilizes probabilistic partial charging, which covers more nodes in each MCV queue. The nodes in each MCV queue are assigned a probabilistic charging factor based on their relative criticality. The charging factor strategy incorporates the charging control factor, assisting in maximizing the reachability of each sensor node in the queue in order to efficiently partially charge the nodes. It also employs the ISAC concept, which revokes multiple MCVs to charge the same node multiple times in a single request, reducing travel costs while increasing the chance of charging the required waiting node in the queue in an efficient manner. 
FLCSD lacked partial charging, while DMCP included it but only focused on relative criticality, neglecting a probabilistic charging factor strategy with a charging control factor that can minimize charging delay.

Fig. \ref{fig: SRNN} depicts the results of survival rate, demonstrating how the results gradually decreases with the number of nodes in the network for all protocols. The proposed protocol \textit{Poised} outperforms cutting-edge protocols with the same reasons given in the charging delay result. 

Fig. \ref{fig: TDNN} depicts the results of travel distance, demonstrating how the results gradually increases with the number of nodes in the network for all protocols. The proposed protocol \textit{Poised} outperforms cutting-edge protocols by utilizing the distance from MCV to charging node distribution in Eq. \eqref{DMC3} which gives priority to nodes in the queue that are closer to MCV. Also, it employs the ISAC approach; for more information, see charging delay result. 
FLCSD and DMCP made no attempt to reduce MCV travel distance in the network.

\section{Conclusion}
This paper presented a probabilistic on-demand charging scheduling for ISAC-assisted WRSNS with multiple mobile charging vehicles. It focuses on the load balanced strategy in charging scheduling for each MCV, the efficient charging factor strategy to partially charge the nodes in the network, and the integrated sensing and communication approach to reduce the traveling cost of each MCV. Therefore, it first considers the four attributes to balance the charging load, along with their probability distributions, in order to prioritize the relatively critical nodes for charging. The attributes are the residual energy of charging node, distance from MCV to charging node, degree of charging node, and charging node betweenness centrality. Second, it focuses on an efficient charging factor strategy to partially charge the nodes based on the charging node residual energy distribution. It also includes a charging control factor to reduce charging delay while increasing charging coverage and survival rate. Finally, the proposed protocol employs the ISAC concept to detect the first arriving MCV within the sensing range of the charging sensor node. This reduces the traveling distance of other MCVs and avoids the conflict of charging the same the node with a single request. 
According to the simulation results, the proposed protocol \textit{Poised} outperforms the cutting-edge protocols.

\section*{Acknowledgment}

This work is supported in part by National Natural Science Foundation of China under Grant 62101232, and in part by the Guangdong Provincial Natural Science Foundation under Grant 2022A1515011257.

\bibliographystyle{ieeetr}
{\footnotesize
	\bibliography{sample-base}}

\end{document}